\documentstyle[times,twocolumn,latex8]{article}

\newcommand{\OR}{\mbox{\rm OR}}

\newtheorem{theorem}{Theorem}
\newtheorem{lemma}{Lemma}

\newcommand{\st}[1]{|#1\rangle} 

\newcommand{\olo}{\vec{0}}
\newcommand{\ceil}[1]{\lceil{#1}\rceil} 
\def\01{\{0,1\}} 

\def\half{\textstyle{1 \over 2}}
\def\s2{{\textstyle{1 \over \sqrt{2}}}}

\def\xor{\oplus}
\def\e{\varepsilon}

\newenvironment{proof}
{\noindent {\bf Proof }}
{{\hfill $\Box$}\\
}

\begin{document}

\title{Bounds for Small-Error and Zero-Error Quantum Algorithms}
\author{
Harry Buhrman\\
{\protect\small\sl CWI\/}%
\thanks{CWI, INS4, P.O.~Box 94079, 1090 GB Amsterdam, The Netherlands.
E-mail: {\tt buhrman@cwi.nl}.}
\and
Richard Cleve\\
{\protect\small\sl University of Calgary\/}%
\thanks{Department of Computer Science, University of Calgary, Calgary, Alberta,
Canada T2N 1N4. E-mail: {\tt cleve@cpsc.ucalgary.ca}.}
\and
Ronald de Wolf\\
{\protect\small\sl CWI and U.~of~Amsterdam\/}%
\thanks{CWI, INS4, P.O.~Box 94079, 1090 GB Amsterdam, The Netherlands.
E-mail: {\tt rdewolf@cwi.nl}.}
\and
Christof Zalka\\
{\protect\small\sl Los Alamos\/}%
\thanks{MS B288, Los Alamos National Laboratory, NM, USA. E-mail: {\tt zalka@t6-serv.lanl.gov}.}
\and
}
\maketitle
\thispagestyle{empty}

\begin{abstract}
We present a number of results related to quantum algorithms with 
small error probability and quantum algorithms that are zero-error.
First, we give a tight analysis of the trade-offs between the number 
of queries of quantum search algorithms, their error probability, 
the size of the search space, and the number of solutions in this 
space.
Using this, we deduce new lower and upper bounds for quantum versions 
of amplification problems.
Next, we establish nearly optimal quantum-classical separations 
for the query complexity of monotone functions in the zero-error 
model (where our quantum zero-error model is defined so as to be 
robust when the quantum gates are noisy).
Also, we present a communication complexity problem related 
to a total function for which there is a quantum-classical communication 
complexity gap in the zero-error model.
Finally, we prove separations for monotone graph properties in the 
zero-error and other error models which imply that the evasiveness 
conjecture for such properties does not hold for quantum computers.
\end{abstract}

\section{Motivation and summary of results}\label{sec:intro}

A general goal in the design of randomized algorithms is to obtain 
fast algorithms with {\em small error probabilities}.
Along these lines is also the goal of obtaining fast algorithms that 
are zero-error (a.k.a.~Las Vegas), as opposed to bounded-error 
(a.k.a.~Monte Carlo).
We examine these themes in the context of {\em quantum\/} 
algorithms, and present a number of new upper and lower bounds that 
contrast with those that arise in the classical case.

The error probabilities of many classical probabilistic algorithms can be 
reduced by techniques that are commonly referred to as {\em amplification}.
For example, if an algorithm $A$ that errs with probability 
$\le {1 \over 3}$ is known, then an error probability bounded above by an 
arbitrarily small $\e>0$ can be obtained by running $A$ independently 
$\Theta(\log({1/\e}))$ times and taking the majority value of the outcomes.
This amplification procedure increases the running time of the algorithm 
by a multiplicative factor of $\Theta(\log({1/\e}))$ and is optimal 
(assuming that $A$ is only used as a black-box).
We first consider the question of whether or not it is possible to 
perform amplification more efficiently on a {\em quantum\/} computer.

A classical probabilistic algorithm $A$ is said to 
{\em $(p,q)$-compute} a function $f : \01^{\ast} \rightarrow \01$ if 
\begin{eqnarray*}
\label{classical}
\Pr[A(x)=1] \!& &\! \cases{ \,\le p & if $f(x) = 0$\cr
                        \,\ge q & if $f(x) = 1$.\cr}
\end{eqnarray*}
Algorithm $A$ can be regarded as a {\em deterministic} algorithm with 
an auxiliary input $r$, which is uniformly distributed over some 
underlying sample space $S$ (usually $S$ is of the form $\01^{l(|x|)}$).
We will focus our attention on the {\em one-sided-error\/} case 
(i.e.\ when $p=0$) and prove bounds on quantum amplification by 
translating them to bounds on quantum search.
In this case, for any $x\in\01^n$, $f(x)=1$ iff 
$(\exists r \in S)(A(x,r)=1)$.

Grover's quantum search algorithm~\cite{grover:search} (and some refinements 
of it~\cite{bbht:bounds,brassard&hoyer:simon,bht:counting,mosca:eigen,
grover:framework}) 
can be cast as a quantum amplification method that is provably more efficient 
than any classical method.
It amplifies a $(0,q)$-algorithm to a 
$(0,\half)$-quantum-computer with $O(1/\sqrt{q})$ executions of $A$, whereas 
classically $\Theta(1/q)$ executions of $A$ would be required to achieve this.
It is natural to consider other amplification problems, such as amplifying 
$(0,q)$-computers to $(0,1 - \e)$-quantum-computers ($0<q<1 - \e<1$).
We give a tight analysis of this.

\begin{theorem}\label{amp}
Let $A : \01^n \times S \rightarrow \{0,1\}$ be a classical probabilistic 
algorithm that $(0,q)$-computes some function $f$, and let $N=|S|$
and $\e\geq 2^{-N}$.
Then, given a black-box for $A$, the number of calls to $A$ 
that are necessary and sufficient to $(0,1-\e)$-quantum-compute $f$ is
\begin{eqnarray}\label{ampbound}
\Theta\left(\sqrt{N}\left({\textstyle{\sqrt{\log(1 / \e) + q N}}} 
- \sqrt{q N}\right)\right).
\end{eqnarray}
\end{theorem}

The lower bound is proven via the polynomial 
method~\cite{nisan&szegedy:degree,bbcmw:polynomials} and with 
adaptations of techniques from~\cite{paturi:degree,coppersmith&rivlin:poly}.
The upper bound is obtained by a combination of ideas, including repeated 
calls to an exact quantum search algorithm for the special case where the 
exact number of solutions is known~\cite{brassard&hoyer:simon,bht:counting}.

{}From Theorem~\ref{amp} we deduce that amplifying $(0,\half)$ 
classical computers to $(0,1-\e)$ quantum computers requires 
$\Theta(\log(1/\e))$ executions, and hence cannot be done 
more efficiently in the quantum case than in the classical case.
These bounds also imply a remarkable algorithm for amplifying a classical 
$(0,{1 \over N})$-computer $A$ to a $(0,1-\e)$ quantum computer.
Note that if we follow the natural approach of composing an optimal 
$(0,{1\over N}) \rightarrow (0,\half)$ amplifier with an optimal 
$(0,\half) \rightarrow (0,1-\e)$ amplifier then our amplifier makes 
$\Theta(\sqrt{N}\log(1 / \e))$ calls to $A$.
On the other hand, Theorem~1 shows that, in the case where $N = |S|$, 
there is a more efficient $(0,{1 \over N}) \rightarrow (0,1-\e)$ 
amplifier that makes only $\Theta(\sqrt{N\log(1/\e)})$ calls to $A$ 
(and this is optimal).

Next we turn our attention to the {\em zero-error\/} (Las Vegas) model.
A zero-error algorithm never outputs an incorrect answer but it may
claim ignorance (output `inconclusive') with probability $\leq 1/2$.
Suppose we want to compute some function $f:\01^N\rightarrow\01$.
The input $x\in\01^N$ can only be accessed by means of queries to 
a black-box which returns the $i$th bit of $x$ when queried on $i$.
Let $D(f)$ denote the number of variables that a {\em deterministic} classical
algorithm needs to query (in the worst case) in order to compute $f$, 
$R_0(f)$ the number of queries for a {\em zero-error} classical algorithm, 
and $R_2(f)$ for {\em bounded-error}.
There is a monotone function $g$ with
$R_0(g)\in O(D(g)^{0.753\ldots})$~\cite{snir:trees,saks&wigderson:trees}, 
and it is known that $R_0(f) \geq \sqrt{D(f)}$ for any function 
$f$~\cite{blum&impagliazzo:generic,hh:oneway}.
It is a longstanding open question whether $R_0(f)\geq\sqrt{D(f)}$ is tight.
We solve the analogous question for monotone functions for the quantum case.

Let $Q_E(f)$, $Q_0(f)$, $Q_2(f)$ respectively be the number of queries that 
an exact, zero-error, or bounded-error quantum algorithm must make to compute $f$.
For zero-error quantum algorithms, there is an issue about the precision 
with which its gates are implemented: any slight imprecisions can reduce 
an implementation of a zero-error algorithm to a bounded-error one.
We address this issue by requiring our zero-error quantum algorithms to 
be {\em self-certifying\/} in the sense that they produce, with constant 
probability, a {\em certificate\/} for the value of $f$ that can be 
verified by a {\em classical\/} algorithm.
As a result, the algorithms remain zero-error even with imperfect quantum 
gates.
The number of queries is then counted as the {\em sum} of those of the 
quantum algorithm (that searches for a certificate) and the classical 
algorithm (that verifies a certificate).
Our upper bounds for $Q_0(f)$ will all be with self-certifying algorithms. 

We first show that $Q_0(f)\geq \sqrt{D(f)}$ for every {\em monotone} $f$
(even without the self-certifying requirement).
Then we exhibit a family of monotone functions that nearly achieves this gap: 
for every $\e>0$ we construct a $g$ such that $Q_0(g)\in O(D(g)^{0.5 + \e})$.
In fact even $Q_0(g)\in O(R_2(g)^{0.5 + \e})$.
These $g$ are so-called ``AND-OR-trees''.
They are the first examples of functions $f:\01^N \rightarrow \01$ 
whose quantum zero-error query complexity is asymptotically less than 
their classical zero-error or bounded-error query complexity.
It should be noted that $Q_0(\mbox{OR})=N$~\cite{bbcmw:polynomials}, 
so the quadratic speedup from Grover's algorithm is lost when zero-error 
performance is required.

Furthermore, we apply the idea behind the above zero-error quantum algorithms 
to obtain a new result in communication complexity.
We derive from the AND-OR-trees a communication complexity problem
where an asymptotic gap occurs between the zero-error quantum communication 
complexity and the zero-error classical communication complexity 
(there was a previous example of a zero-error gap for a function with 
restricted domain in~\cite{BuhrmanCleveWigderson98} and bounded-error gaps 
in~\cite{astu:qsampling,raz:qcc}).
This result includes a new lower bound in classical communication complexity.
We also state a result by Hartmut Klauck, inspired by an earlier version of
this paper, which gives the first {\em total} function with 
quantum-classical gap in the zero-error model of communication complexity.

Finally, a class of black-box problems that has received wide attention 
concerns the determination of monotone graph properties
\cite{rivest&vuillemin,kss:topo,king:graphprop,hajnal:graphprop}.
Consider a directed graph on $n$ vertices.
It has $n(n-1)$ possible edges and hence can be represented
by a black-box of $n(n-1)$ binary variables, where
each variable indicates whether or not a specific edge is present.
A {\em nontrivial monotone graph property\/} is a property of such a 
graph (i.e.~a function $P : \{0,1\}^{n(n-1)} \rightarrow \{0,1\}$) 
that is non-constant, invariant under permutations of the vertices 
of the graph, and monotone.
Clearly, $n(n-1)$ is an upper bound on the number of queries required to 
compute such properties.
The {\em Aanderaa-Karp-Rosenberg} or {\em evasiveness conjecture}
states that $D(P)=n(n-1)$ for all $P$. The best known general lower bound
is $\Omega(n^2)$~\cite{rivest&vuillemin,kss:topo,king:graphprop}.
It has also been conjectured that $R_0(P)\in\Omega(n^2)$ for all $P$,
but the current best bound is only $\Omega(n^{4/3})$~\cite{hajnal:graphprop}.
A natural question is whether or not quantum algorithms can determine
monotone graph properties more efficiently.
We show that they can.
Firstly, in the exact model we exhibit a $P$ with $Q_E(P)<n(n-1)$, 
so the evasiveness conjecture fails in the case of quantum computers.
However, we also prove $Q_E(P)\in\Omega(n^2)$ for all $P$, so evasiveness
does hold up to a constant factor for exact quantum computers.
Secondly, we give a nontrivial monotone graph property for which the 
evasiveness conjecture is violated by a zero-error quantum algorithm:
let STAR be the property that the graph 
has a vertex which is adjacent to all other vertices.
Any classical (zero-error or bounded-error) algorithm for STAR requires 
$\Omega(n^2)$ queries.  We give a zero-error quantum algorithm that 
determines STAR with only $O(n^{3/2})$ queries.
Finally, for bounded-error quantum algorithms, the OR problem
trivially translates into the monotone graph property
``there is at least one edge'', which can be determined with only $O(n)$
queries via Grover's algorithm~\cite{grover:search}.

\section{Basic definitions and terminology}

See~\cite{berthiaume:qc,bbcmw:polynomials} for details and references
for the quantum circuit model.
For $b\in\01$, a query gate $O$ for an input $x=(x_0,\ldots,x_{N-1})\in\01^N$ 
performs the following mapping, which is our only way to access the bits $x_j$:
$$
\st{j,b}\rightarrow\st{j,b\xor x_j}.
$$
We sometimes use the term ``black-box'' for $x$ as well as $O$.
A quantum algorithm or gate network $A$ with $T$ queries is 
a unitary transformation $A=U_TOU_{T-1}O\ldots OU_1OU_0$.
Here the $U_i$ are unitary transformations that do not depend on $x$.
Without loss of generality we fix the initial state to $\st{\olo}$,
independent of $x$.
The final state is then a superposition $A\st{\olo}$
which depends on $x$ only via the $T$ query gates.
One specific qubit of the final state (the rightmost one, say) 
is designated for the output.
The acceptance probability of a quantum network on a specific 
black-box $x$ is defined to be the probability that the output qubit 
is 1 (if a measurement is performed on the final state).

We want to compute a function $f:\01^N\rightarrow\01$, 
using as few queries as possible (on the worst-case input).
We distinguish between three different error-models. 
In the case of {\em exact} computation, an algorithm must always give 
the correct answer $f(x)$ for every $x$.
In the case of {\em bounded-error} computation, an algorithm must give 
the correct answer $f(x)$ with probability $\geq 2/3$ for every $x$.
In the case of {\em zero-error} computation, an algorithm is allowed
to give the answer `don't know' with probability $\leq 1/2$, but {\em if}
it outputs an answer (0 or 1), then this must be the correct answer.
The complexity in this zero-error model is equal up to a factor of 2
to the {\em expected} complexity of an optimal algorithm that always
outputs the correct answer.
Let $D(f)$, $R_0(f)$, and $R_2(f)$ denote the exact, zero-error and
bounded-error classical complexities, respectively, and 
$Q_E(f)$, $Q_0(f)$, $Q_2(f)$ be the corresponding quantum complexities.
Note that $N\geq D(f)\geq Q_E(f)\geq Q_0(f)\geq Q_2(f)$ and
$N\geq D(f)\geq R_0(f)\geq R_2(f)\geq Q_2(f)$ for every $f$.

\section{Tight trade-offs for quantum searching}\label{secerrorbound}

In this section, we prove Theorem~\ref{amp}, stated in 
Section~\ref{sec:intro}.
The {\em search} problem is the following: for a given black-box $x$, 
find a $j$ such that $x_j=1$ using as few queries to $x$ as possible.  
A quantum computer can achieve error probability 
$\leq 1/3$ using $T\in\Theta(\sqrt{N})$ 
queries~\cite{grover:search}.  We address the question of how large the
number of queries should be in order to be able to achieve a very small error
$\e$.  We will prove that if $T<N$, then $
T\in\Theta\left(\sqrt{N\log(1/\e)}\right).  $ This result will actually be a
special case of a more general theorem that involves a promise on the number of
solutions.  Suppose we want to search a space of $N$ items with error $\e$, and
we are promised that there are  at least some number $t<N$
solutions.  The higher $t$ is, the fewer queries we will need.  In the appendix
we give the following lower bound on $\e$ in terms of $T$, using tools
from~\cite{bbcmw:polynomials,paturi:degree,coppersmith&rivlin:poly}.

\begin{theorem}\label{thqerrorboundt}
Under the promise that the number of solutions is at least $t$,
every quantum search algorithm that uses $T\leq N-t$ queries 
has error probability
$$
\e\in\Omega\left(e^{-4bT^2/(N-t)-8T\sqrt{tN/(N-t)^2}}\right).
$$
\end{theorem}

Here $b$ is a positive universal constant.
This theorem implies a lower bound on $T$ in terms of $\e$.
To give a tight characterization of the relations between $T$, $N$, $t$ 
and $\e$, we need the following upper bound on $T$ for the case $t=1$:

\begin{theorem}\label{thupperboundT}
For every $\e>0$ there exists a quantum search algorithm
with error probability $\leq\e$ and
$\displaystyle O\left(\sqrt{N\log(1/\e)}\right)$ queries.
\end{theorem}

\begin{proof}
Set $t_0=\ceil{\log(1/\e)}$. Consider the following algorithm:
\begin{enumerate}
\item Apply exact search for $t=1,\ldots,t_0$, each of which takes
$O(\sqrt{N/t})$ queries.
\item If no solution has been found, then conduct $t_0$
searches, each with $O(\sqrt{N/t_0})$ queries.
\item Output a solution if one has been found, otherwise output `no'.
\end{enumerate}
The query complexity of this algorithm is bounded by
$$
\sum_{t=1}^{t_0}O\left(\sqrt{\frac{N}{t}}\right)+
t_0 O\left(\sqrt{\frac{N}{t_0}}\right)
=O\left(\sqrt{N\log(1/\e)}\right).
$$
If the real number of solutions was in $\{1,\ldots,t_0\}$, 
then a solution will be found with certainty in step~1.
If the real number of solutions was $>t_0$, then each of the searches 
in step~2 can be made to have error probability $\leq 1/2$, 
so we have total error probability at most $(1/2)^{t_0}\leq\e$.
\end{proof}

A more precise analysis gives $T\leq 2.45\sqrt{N\log(1/\e)}$. 
It is interesting that we can use this to prove something about
the constant $b$ of the Coppersmith-Rivlin theorem (see appendix):
for $t=1$ and $\e\in o(1)$, the lower bound
asymptotically becomes $T\geq\sqrt{N\log(1/\e)/4b}$.
Together these two bounds imply $b\geq 1/4(2.45)^2\approx 0.042$.

The main theorem of this section tightly characterizes the various 
trade-offs between the size of the search space $N$, the promise $t$, 
the error probability $\e$, and the required number of queries:

\begin{theorem}\label{thm:tight}
Fix $\eta\in(0,1)$, and let $N>0$, $\e\geq 2^{-N}$, and $t\leq\eta N$.
Let $T$ be the optimal number of queries a quantum computer
needs to search with error $\leq\e$ through an unordered list of $N$ items
containing at least $t$ solutions.
Then
$$
\log(1/\e)\in\Theta\left(\frac{T^2}{N}+T\sqrt{\frac{t}{N}}\right).
$$
\end{theorem}

\begin{proof}
{}From Theorem~\ref{thqerrorboundt} we obtain the upper bound 
$\displaystyle
\log(1/\e)\in O\left(\frac{T^2}{N}+T\sqrt{\frac{t}{N}}\right).$
To prove a lower bound on $\log(1/\e)$ we distinguish two cases.

{\bf Case 1: $T\geq\sqrt{tN}$.}
By Theorem~\ref{thupperboundT}, we can achieve error $\leq\e$
using $T_u\in O(\sqrt{N\log(1/\e)})$ queries. 
Now (leaving out some constant factors):
$$
\log(1/\e)\geq\frac{T_u^2}{N}\geq
\frac{1}{2}\left(\frac{T^2}{N}+T\frac{T}{N}\right)\geq
\frac{1}{2}\left(\frac{T^2}{N}+T\sqrt{\frac{t}{N}}\right).
$$

{\bf Case 2: $T<\sqrt{tN}$.}
We can achieve error $\leq 1/2$ using $O(\sqrt{N/t})$ queries, 
and then classically amplify this to error $\leq 1/\e$
using $O(\log(1/\e))$ repetitions.
This takes $T_u\in O(\sqrt{N/t}\log(1/\e))$ queries in total.
Now:
$$
\log(1/\e)\geq T_u\sqrt{\frac{t}{N}}\geq
\frac{1}{2}\left(T\sqrt{\frac{t}{N}}+T\sqrt{\frac{t}{N}}\right)\geq
$$
$$
\frac{1}{2}\left(\frac{T^2}{N}+T\sqrt{\frac{t}{N}}\right).
$$

\vspace*{-7mm}

\end{proof}

Rewriting Theorem~\ref{thm:tight} (with $q=t/N$) yields the general 
bound of Theorem~\ref{amp}.

For $t=1$ this becomes $T\in\Theta(\sqrt{N\log(1/\e)})$.
Thus no quantum search algorithm with $O(\sqrt{N})$ 
queries has error probability $o(1)$.
Also, a quantum search algorithm with $\e\leq 2^{-N}$
needs $\Omega(N)$ queries.
For the case $\e=1/3$ we re-derive the bound $\Theta(\sqrt{N/t})$ 
from~\cite{bbht:bounds}.

\section{Applications of Theorem 1 to amplification}\label{secnoampl}

In this section we apply the bounds from Theorem~\ref{amp} to examine
the speedup possible for amplifying classical one-sided error 
algorithms via quantum algorithms.
Observe that searching for items in a search space of size $N$ and
figuring out whether a probabilistic one-sided error algorithm $A$ with 
sample space $S$ of size $N$ accepts are essentially the same thing.

Let us analyze some special cases more closely.
Suppose that we want to amplify an algorithm $A$ that $(0,\half)$-computes 
some function $f$ to $(0,1-\e)$.
Then substituting $|S| = N$ and $q = \half$ into Eq.~(\ref{ampbound}) 
in Theorem~\ref{amp} yields

\begin{theorem}\label{thnoamplrp}
Let $A : \01^n \times S \rightarrow \{0,1\}$ be a classical probabilistic 
algorithm that $(0,\half)$-computes some function $f$, and $\e\geq 2^{-|S|}$.
Then, given a black-box for $A$, the number of calls to $A$ 
that any quantum algorithm needs to make to $(0,1-\e)$-compute $f$ is
$\Omega(\log(1/\e))$.
\end{theorem}

Hence amplification of one-sided error algorithms with fixed initial
success probability cannot be done more efficiently in the quantum case 
than in the classical case.
Since one-sided error algorithms are a special case of
bounded-error algorithms, the same lower bound also holds for amplification 
of bounded-error algorithms.
A similar but slightly more elaborate argument as above shows that
a quantum computer still needs $\Omega(\log(1/\e))$ applications of $A$
when $A$ is zero-error.


Some other special cases of Theorem~\ref{amp}:
in order to amplify a $(0,{1 \over N})$-computer $A$ to a 
$(0,\half)$-computer, $\Theta(\sqrt{N})$ calls to $A$ are necessary 
and sufficient (and this is essentially a restatement of known results 
of Grover and others about quantum searching~\cite{grover:search,bbht:bounds}).
Also, in order to amplify a $(0,{1 \over N})$-computer with sample 
space of size $N$ to a $(0,1 - \e)$-computer, 
$\Theta(\sqrt{N \log(1 / \e)})$ calls to $A$ are necessary and sufficient.

Finally, consider what happens if the size of the sample space is unknown
and we only know that $A$ is a classical one-sided error algorithm 
with success probability $q$.
Quantum amplitude amplification can improve the success probability to
$1/2$ using $O(1/\sqrt{q})$ repetitions of $A$. We can then classically
amplify the success probability further to $1-\e$ using $O(\log(1/\e))$
repetitions. In all, this method uses $O(\log(1/\e)/\sqrt{q})$ applications
of $A$. Theorem~\ref{thm:tight} implies that this is best possible 
in the worst case (i.e.~if $A$ happens to be a classical algorithm
with very large sample space).

\section{Zero-error quantum algorithms}\label{secsepzero}

In this section we consider zero-error complexity of functions
in the query (a.k.a. black-box) setting. The best general bound that 
we can prove between the quantum zero-error complexity $Q_0(f)$ and the 
classical deterministic complexity $D(f)$ for total functions 
is the following (the proof is similar to the $D(f)\in O(Q_E(f)^4)$ 
result given in~\cite{bbcmw:polynomials} and uses an unpublished proof 
technique of Nisan and Smolensky):

\begin{theorem}
For every total function $f$ we have $D(f)\in O(Q_0(f)^4)$.
\end{theorem}

We will in particular look at {\em monotone increasing} $f$. 
Here the value of $f$ cannot flip from 1 to 0 if more variables are set to 1.
For such $f$, we improve the bound to:

\begin{theorem}\label{thdvsqo}
For every total monotone Boolean function $f$ we have
$D(f)\leq Q_0(f)^2$.
\end{theorem}

\begin{proof}
Let $s(f)$ be the {\em sensitivity} of $f$: the maximum, over all $x$, of
the number of variables that we can individually flip in $x$ to change $f(x)$.
Let $x$ be an input on which the sensitivity of $f$ equals $s(f)$.
Assume without loss of generality that $f(x)=0$.
All sensitive variables must be 0 in $x$, and setting one or more of
them to 1 changes the value of $f$ from 0 to 1.
Hence by fixing all variables in $x$ except for the $s(f)$
sensitive variables, we obtain the OR function on $s(f)$ variables.
Since OR on $s(f)$ variables has 
$Q_0(\OR)=s(f)$~\cite[Proposition~6.1]{bbcmw:polynomials},
it follows that $s(f)\leq Q_0(f)$.
It is known (see for instance~\cite{nisan:pram&dt,bbcmw:polynomials}) 
that $D(f)\leq s(f)^2$ for monotone $f$, hence $D(f)\leq Q_0(f)^2$.
\end{proof}

Important examples of monotone functions are {\em AND-OR trees}. 
These can be represented as trees of depth $d$ where the $N$ leaves 
are the variables, and the $d$ levels of internal nodes are alternatingly 
labeled with ANDs and ORs. Using techniques from~\cite{bbcmw:polynomials},
it is easy to show that $Q_E(f)\geq N/2$ and $D(f)=N$ for such trees.
However, we show that in the zero-error setting quantum computers can achieve
significant speed-ups for such functions.
These are in fact the first total functions with superlinear gap
between quantum and classical zero-error complexity.
Interestingly, the quantum algorithms for these functions are not just
zero-error: if they output an answer $b\in\01$ then they also output 
a {\em $b$-certificate} for this answer. This is a set of indices 
of variables whose values force the function to the value $b$.

We prove that for sufficiently large $d$, quantum computers 
can obtain near-quadratic speed-ups on $d$-level AND-OR trees which 
are uniform, i.e.~have branching factor $N^{1/d}$ at each level.
Using the next lemma (which is proved in the appendix) we show that 
Theorem~\ref{thdvsqo} is almost tight:
for every $\e>0$ there exists a total monotone $f$ with 
$Q_0(f)\in O(N^{1/2+\e})$.

\begin{lemma}\label{lemandord}
Let $d\geq 1$ and let $f$ denote the uniform $d$-level AND-OR tree 
on $N$ variables that has an OR as root. 
There exists a quantum algorithm $A_1$ that finds a 1-certificate 
in expected number of queries $O(N^{1/2+1/2d})$ if $f(x)=1$ and 
does not terminate if $f(x)=0$.
Similarly, there exists a quantum algorithm $A_0$ that finds a 0-certificate 
in expected number of queries $O(N^{1/2+1/d})$ if $f(x)=0$ and 
does not terminate if $f(x)=1$.
\end{lemma}

\begin{theorem}\label{thmonsep}
Let $d\geq 1$ and let $f$ denote the uniform $d$-level AND-OR tree 
on $N$ variables that has an OR as root. 
Then $Q_0(f)\in O(N^{1/2+1/d})$ and $R_2(f)\in\Omega(N)$.
\end{theorem}

\begin{proof}
Run the algorithms $A_1$ and $A_0$ of Lemma~\ref{lemandord}
side-by-side until one of them terminates with a certificate.
This gives a certificate-finding quantum algorithm for $f$
with expected number of queries $O(N^{1/2+1/d})$.
Run this algorithm for twice its expected number of queries and
answer `don't know' if it hasn't terminated after that time. 
By Markov's inequality, the probability of non-termination is $\leq 1/2$,
so we obtain an algorithm for our zero-error setting with
$Q_0(f)\in O(N^{1/2+1/d})$ queries.

The classical lower bound follows from combining two known results.
First, an AND-OR tree of depth $d$ on $N$ variables has $R_0(f)\geq N/2^d$
\cite[Theorem~2.1]{hnw:readonce} (see also~\cite{saks&wigderson:trees}). 
Second, for such trees we have $R_2(f)\in\Omega(R_0(f))$~\cite{santha:montecarlo}.
Hence $R_2(f)\in\Omega(N)$.
\end{proof}

This analysis is not quite optimal.
It gives only trivial bounds for $d=2$, but a more refined 
analysis shows that we can also get speed-ups for such 2-level trees:

\begin{theorem}\label{thgapqo}
Let $f$ be the AND of $N^{1/3}$ ORs of $N^{2/3}$ variables each.
Then $Q_0(f)\in \Theta(N^{2/3})$ and $R_2(f)\in\Omega(N)$.
\end{theorem}

\begin{proof}
A similar analysis as before shows $Q_0(f)\in O(N^{2/3})$ and
$R_2(f)\in\Omega(N)$.

For the quantum lower bound: note that if we set all variables to 1 
except for the $N^{2/3}$ variables in the first subtree, then $f$ becomes
the OR of $N^{2/3}$ variables. This is known to have zero-error complexity
exactly $N^{2/3}$~\cite[Proposition~6.1]{bbcmw:polynomials}, 
hence $Q_0(f)\in\Omega(N^{2/3})$.
\end{proof}

If we consider a tree with $\sqrt{N}$ 
subtrees of $\sqrt{N}$ variables each, we would get $Q_0(f)\in O(N^{3/4})$
and $R_2(f)\in\Omega(N)$. The best lower bound we can prove here is 
$Q_0(f)\in\Omega(\sqrt{N})$. However, if we also require the quantum
algorithm to output a {\em certificate} for $f$, we can prove a tight
quantum lower bound of $\Omega(N^{3/4})$. We do not give the proof here,
which is a technical and more elaborate version of the proof of
the classical lower bound of Theorem~\ref{thcertgap}.

\section{Zero-error communication complexity}

The results of the previous section can be translated to the
setting of communication complexity~\cite{kushilevitz&nisan:cc}. 
Here there are two parties, Alice and Bob, who want 
to compute some relation $R\subseteq\01^N\times\01^N\times\01^M$. 
Alice gets input $x\in\01^N$ and Bob gets input $y\in\01^N$. 
Together they want to compute some $z\in\01^M$ such that $(x,y,z)\in R$,
exchanging as few bits of communication as possible.
The often studied setting where Alice and Bob want to compute some 
function $f:\01^N\times\01^N\rightarrow\01$ is a special case of this.
In the case of {\em quantum} communication, Alice and Bob can exchange and
process qubits, potentially giving them more power than classical 
communication.

Let $g:\01^N\rightarrow\01$ be one of the AND-OR-trees of the previous section.
We can derive from this a communication problem
$f:\01^N\times\01^N\rightarrow\01$ 
by defining $f(x,y)=g(x\wedge y)$, where $x\wedge y\in\01^N$ is the vector 
obtained by bitwise AND-ing Alice's $x$ and Bob's $y$.
Let us call such a problem a ``distributed'' AND-OR-tree.
Buhrman, Cleve, and Wigderson~\cite{BuhrmanCleveWigderson98} 
show how to turn a $T$-query quantum black-box algorithm for $g$ into 
a communication protocol for $f$ with $O(T\log N)$ qubits of communication.
Thus, using the upper bounds of the previous section, for every $\e>0$, 
there exists a distributed AND-OR-tree $f$ that has a $O(N^{1/2+\e})$-qubit
zero-error protocol.
It is conceivable that the classical zero-error communication complexity 
of these functions is $\omega(N^{1/2+\e})$; however, we are not able 
to prove such a lower bound at this time.
Nevertheless, we are able to establish a quantum-classical separation for 
a relation that is closely related to the AND-OR-tree functions, 
which is explained below.

For any AND-OR tree function $g : \01^N \rightarrow \01$ and input 
$x \in \01^N$, a {\em certificate for the value of $g$ on input $x$} 
is a subset $c$ of the indices $\{0,1,\ldots,N-1\}$ such that the 
values $\{x_i : i \in c\}$ determine the value of $g(x)$.
It is natural to denote $c$ as an element of $\01^N$, representing 
the characteristic function of the set.
For example, for 
\begin{equation}
\label{cert_example}
g(x_0,x_1,x_2,x_3) = (x_0 \vee x_1) \wedge (x_2 \vee x_3),
\end{equation} 
a certificate for the value of $g$ on input $x = 1011$ is $c = 1001$, 
which indicates that $x_0 = 1$ and $x_3 = 1$ determine the value of $g$.

We can define a communication problem based on finding these certificates 
as follows.
For any AND-OR tree function $g : \01^N \rightarrow \01$ and 
$x, y \in \01^N$, a certificate for the value of $g$ on {\em distributed\/} 
inputs $x$ and $y$ is a subset $c$ of $\{0,1,\ldots,N-1\}$ (denoted as 
an element of $\01^N$) such that the 
values $\{(x_i,y_i) : i \in c\}$ determine the value of $g(x \wedge y)$.
Define the relation 
$R\subseteq\01^N\times\01^N\times\01^N$ 
such that $(x,y,c)\in R$ iff $c$ is a certificate for the value 
of $g$ on distributed inputs $x$ and $y$.
For example, when $R$ is with respect to the function $g$ of 
equation~(\ref{cert_example}), $(1011,1111,1001) \in R$, because, 
for $x=1011$ and $y=1111$, an appropriate certificate is $c=1001$.

The zero-error certificate-finding algorithm for $g$ of the 
previous section, together with the 
\cite{BuhrmanCleveWigderson98}-translation 
from black-box algorithms to communication protocols, 
implies a zero-error quantum communication protocol for $R$.
Thus, Theorem~\ref{thmonsep} implies that for every $\varepsilon>0$ 
there exists a relation $R\subseteq\01^N\times\01^N\times\01^N$ 
for which there is a zero-error quantum protocol with $O(N^{1/2+\e})$ 
qubits of communication.
Although we suspect that the {\em classical} zero-error communication 
complexity of these relations is $\Omega(N)$, we are only able to 
prove lower bounds for relations derived from 2-level trees:

\begin{theorem}\label{thcertgap}
Let $g:\01^N\rightarrow\01$ be an AND of $N^{1/3}$ ORs of $N^{2/3}$ 
variables each.
Let $R\subseteq\01^N\times\01^N\times\01^N$ be 
the certificate-relation derived from $g$.
Then there exists a zero-error $O(N^{2/3}\log N)$-qubit quantum 
protocol for $R$, whereas, any zero-error classical protocol for 
$R$ needs $\Omega(N)$ bits of communication.
\end{theorem}

\begin{proof}
The quantum upper bound follows from Theorem~\ref{thgapqo} and the
\cite{BuhrmanCleveWigderson98}-reduction.

For the classical lower bound, suppose we have a classical zero-error 
protocol $P$ for $R$ with $T$ bits of communication.
We will show how we can use this to solve the Disjointness problem
on $k=N^{1/3}(N^{2/3}-1)$ variables. 
(Given Alice's input $x\in\01^k$ and Bob's $y\in\01^k$, the Disjointness 
problem 
is to determine if $x$ and $y$ have a 1 at the same position somewhere.)
Let $Q$ be the following classical protocol. 
Alice and Bob view their $k$-bit input as made up of $N^{1/3}$
subtrees of $N^{2/3}-1$ variables each.
They add a dummy variable with value 1 to each subtree and apply 
a random permutation to each subtree (Alice and Bob have to apply
the {\em same} permutation to a subtree, so we assume a public coin). 
Call the $N$-bit strings they now have $x'$ and $y'$.
Then they apply $P$ to $x'$ and $y'$. 
Since $f(x',y')=1$, after an expected number of $O(T)$ bits of communication
$P$ will deliver a certificate which is a common 1 in each subtree.
If one of these common 1s is non-dummy then Alice and Bob output 1, 
otherwise they output 0.
It is easy to see that this protocol solves Disjointness with success
probability 1 if $x\wedge y=\vec{0}$ and with success probability $\geq 1/2$
if $x\wedge y\neq\vec{0}$. 
It assumes a public coin and uses $O(T)$ bits of communication.
Now the well-known $\Omega(k)$ bound for classical bounded-error Disjointness 
on $k$ variables~\cite{ks:disj,razborov:disj} implies 
$T\in\Omega(k)=\Omega(N)$.
\end{proof}

The relation of Theorem~\ref{thcertgap} is ``total'', in the sense that, 
for every $x, y \in \01^N$, there exists a $c$ such that $(x,y,c) \in R$.
It should be noted that one can trivially construct a total relation from 
any partial function by allowing any output for inputs that are outside 
the domain of the function.
In this manner, a total relation with an exponential quantum-classical 
zero-error gap can be immediately obtained from the distributed 
Deutsch-Jozsa problem of~\cite{BuhrmanCleveWigderson98}.
The total relation of Theorem~\ref{thcertgap} is different from this 
in that it is not a trivial extension of a partial function.

After reading a first version of this paper, Hartmut Klauck proved
a separation which is the first example of a total {\em function\/} 
with superlinear gap between quantum and classical zero-error 
communication complexity~\cite{klauck:qpcom}.
Consider the iterated non-disjointness function: Alice and Bob each 
receive $s$ sets of size $n$ from a size-$poly(n)$ universe (so the 
input length is $N\in\Theta(sn\log n)$ bits), and they have to 
output 1 iff all $s$ pairs of sets intersect. 
Klauck's function $f$ is an intricate subset of this iterated 
non-disjointness function, but still an explicit and total function.
Results of~\cite{hromkovic&schnitger:limadvice} about limited 
non-deterministic communication complexity imply a lower bound for 
classical zero-error protocols for $f$.
On the other hand, because $f$ can be written as a 2-level AND-OR-tree, 
the methods of this paper imply a more efficient quantum zero-error protocol.
Choosing $s=n^{5/6}$, Klauck obtains a polynomial gap:

\begin{theorem}[Klauck \cite{klauck:qpcom}]\label{thklauck}
For $N\in\Theta(n^{11/6}\log n)$ there exists a total function 
$f:\01^N\times\01^N\rightarrow\01$, such that there is a quantum
zero-error protocol for $f$ with $O(N^{10/11+\e})$ qubits of communication
(for all $\e>0$), whereas every classical zero-error protocol for $f$ needs 
$\Omega(N/\log N)$ bits of communication.
\end{theorem}

%

\section{Quantum complexity of graph properties}\label{secgraphprop}

{\em Graph properties} form an interesting subset of the set of
all Boolean functions. Here an input of $N=n(n-1)$ bits
represents the edges of a directed graph on $n$ vertices.
(Our results hold for properties of {\em directed} 
as well as {\em undirected} graphs.)
A {\em graph property} $P$ is a subset of the set of all graphs
that is closed under permutation of the nodes (so if $X,Y$ represent
isomorphic graphs, then $X\in P$ iff $Y\in P$).
We are interested in the number of queries of the form ``is there an
edge from node $i$ to node $j$?'' that we need to determine for 
a given graph whether it has a certain property $P$. 
Since we can view $P$ as a total function on $N$ variables,
we can use the notations $D(P)$, etc.
A property $P$ is {\em evasive} if $D(P)=n(n-1)$, so 
if in the worst case all $N$ edges have to be examined.

The complexity of graph properties has been well-studied classically, 
especially for {\em monotone} graph properties (a property is monotone 
if adding edges cannot destroy the property).
In the sequel, let $P$ stand for a (non-constant) monotone graph property. 
Much research revolved around the so-called Aanderaa-Karp-Rosenberg conjecture 
or {\em evasiveness conjecture}, which states that every $P$ is evasive.
This conjecture is still open; see~\cite{lovasz:graphprop} for an overview. 
It has been proved for $n$ equals a prime power~\cite{kss:topo}, 
but for other $n$ the best known general bound 
is $D(P)\in\Omega(n^2)$ \cite{rivest&vuillemin,kss:topo,king:graphprop}.
(Evasiveness has also been proved for {\em bipartite} graphs~\cite{yao:bipartite}.)
For the classical zero-error complexity, the best known general
result is $R_0(P)\in\Omega(n^{4/3})$~\cite{hajnal:graphprop},
but it has been conjectured that $R_0(P)\in\Theta(n^2)$. 
To the best of our knowledge, no $P$ is known to have $R_2(P)\in o(n^2)$.

In this section we examine the complexity of monotone graph properties 
on a quantum computer.
First we show that if we replace exact classical algorithms by exact
quantum algorithms, then the evasiveness conjecture fails.
However, the conjecture does hold up to a constant factor.

\begin{theorem}
For all $P$, $Q_E(P)\in\Omega(n^2)$.
There is a $P$ such that $Q_E(P)<n(n-1)$ for every $n>2$.
\end{theorem}

\begin{proof}
For the lower bound, let $deg(f)$ denote the degree of the unique
multilinear multivariate polynomial $p$ that represents a function
$f$ (i.e.~$p(X)=f(X)$ for all $X$). 
\cite{bbcmw:polynomials} proves that $Q_E(f)\geq deg(f)/2$ for every $f$. 
Dodis and Khanna~\cite[Theorem~5.1]{dodis&khanna} prove that 
$deg(P)\in\Omega(n^2)$ for all monotone graph properties $P$. 
Combining these two facts gives the lower bound.

Let $P$ be the property ``the graph contains more than $n(n-1)/2$
edges''. This is just a special case of the Majority function.
Let $f$ be Majority on $N$ variables.
It is known that $Q_E(f)\leq N+1-e(N)$, where $e(N)$ is the number of 1s
in the binary expansion of $N$. This was first noted by Hayes, Kutin and Van
Melkebeek~\cite{hkm:qmajority}. It also follows immediately from classical
results~\cite{saks&werman:majority,ars:majority} that show that an item 
with the Majority value can be identified classically deterministically 
with $N-e(N)$ {\em comparisons} between bits (a comparison between two 
black-box-bits is the XOR of two bits, which can be computed with
1 quantum query~\cite{cemm:revisited}).
One further query to this item suffices to determine the Majority value.
For $N=n(n-1)$ and $n>2$ we have $e(N)\geq 2$ and hence 
$Q_E(f)\leq N-e(N)+1<N$.
\end{proof}

In the zero-error case, we can show polynomial gaps between quantum and
classical complexities, so here the evasiveness conjecture fails even
if we ignore constant factors.

\begin{theorem}
For all $P$, $Q_0(P)\in\Omega(n)$.
There is a $P$ such that $Q_0(P)\in O(n^{3/2})$ and $R_2(P)\in\Omega(n^2)$.
\end{theorem}

\begin{proof}
The quantum lower bound follows from $D(P)\leq Q_0(P)^2$ 
(Theorem~\ref{thdvsqo}) and $D(P)\in\Omega(n^2)$.

Consider the property ``the graph contains a star'', where a star is
a node that has edges to all other nodes. This property corresponds to 
a 2-level tree, where the first level is an OR of $n$ subtrees, and each
subtree is an AND of $n-1$ variables. The $n-1$ variables in the $i$th 
subtree correspond to the $n-1$ edges $(i,j)$ for $j\neq i$.
The $i$th subtree is 1 iff the $i$th node is the center of a star, so the
root of the tree is 1 iff the graph contains a star.
Now we can show $Q_0(P)\in O(n^{3/2})$ and $R_2(P)\in\Omega(n^2)$
analogously to Theorem~\ref{thgapqo}.
\end{proof}

Combined with the translation of a quantum algorithm to a polynomial
\cite{bbcmw:polynomials}, this theorem shows that a ``zero-error polynomial''
for the STAR-graph property can have degree $O(n^{3/2})$.
Thus proving a general lower bound on zero-error polynomials for
graph properties will not improve Hajnal's randomized lower bound of 
$n^{4/3}$ further then $n^{3/2}$.
In particular, a proof that $R_0(P)\in\Omega(n^2)$ cannot be obtained
via a lower bound on degrees of polynomials. This contrasts with the case
of exact computation, where the $\Omega(n^2)$ lower bound on $deg(P)$   
implies both $D(P)\in\Omega(n^2)$ and $Q_E(P)\in\Omega(n^2)$.

Finally, for the bounded-error case we have quadratic gaps between
quantum and classical: the property ``the graph has at least one edge''
has $Q_2(P)\in O(n)$ by Grover's quantum search algorithm.
Combining that $D(P)\in\Omega(n^2)$ for all $P$ and
$D(f)\in O(Q_2(f)^4)$ for all monotone $f$~\cite{bbcmw:polynomials},
we also obtain a general lower bound:

\begin{theorem}
For all $P$, we have $Q_2(P)\in\Omega(\sqrt{n})$.
There is a $P$ such that $Q_2(P)\in O(n)$.
\end{theorem}


\noindent
{\large\bf Acknowledgments}\\
We thank Hartmut Klauck for informing us about Theorem~\ref{thklauck}, 
Ramamohan Paturi for Lemma~\ref{lemchebbound}, 
David Deutsch, Wim van Dam, and Michele Mosca for helpful discussions 
which emphasized the importance of small-error quantum search,
and Mosca and Yevgeniy Dodis for helpful discussions about graph properties.


\begin{thebibliography}{10}\setlength{\itemsep}{-1ex}\small

\bibitem{ars:majority}
L.~Alonso, E.~M. Reingold, and R.~Schott.
\newblock Determining the majority.
\newblock {\em Information Processing Letters}, 47(5):253--255, 1993.

\bibitem{astu:qsampling}
A.~Ambainis, L.~Schulman, A.~{Ta-Shma}, U.~Vazirani, and A.~Wigderson.
\newblock The quantum communication complexity of sampling.
\newblock In {\em Proceedings of 39th FOCS}, pages 342--351, 1998.

\bibitem{bbcmw:polynomials}
R.~Beals, H.~Buhrman, R.~Cleve, M.~Mosca, and R.~d. Wolf.
\newblock Quantum lower bounds by polynomials.
\newblock In {\em Proceedings of 39th FOCS}, pages 352--361, 1998.
\newblock quant-ph/9802049.

\bibitem{berthiaume:qc}
A.~Berthiaume.
\newblock Quantum computation.
\newblock In A.~Selman and L.~Hemaspaandra, editors, {\em Complexity Theory
  Retrospective II}, pages 23--51. Springer, 1997.

\bibitem{blum&impagliazzo:generic}
M.~Blum and R.~Impagliazzo.
\newblock Generic oracles and oracle classes (extended abstract).
\newblock In {\em Proceedings of 28th FOCS}, pages 118--126, 1987.

\bibitem{bbht:bounds}
M.~Boyer, G.~Brassard, P.~H{\o}yer, and A.~Tapp.
\newblock Tight bounds on quantum searching.
\newblock {\em Fortschritte der Physik}, 46(4--5):493--505, 1998.
\newblock Earlier version in Physcomp'96. quant-ph/9605034.

\bibitem{brassard&hoyer:simon}
G.~Brassard and P.~H{\o}yer.
\newblock An exact quantum polynomial-time algorithm for {S}imon's problem.
\newblock In {\em Proceedings of the 5th Israeli Symposium on Theory of
  Computing and Systems (ISTCS'97)}, pages 12--23, 1997.
\newblock quant-ph/9704027.

\bibitem{bht:counting}
G.~Brassard, P.~H{\o}yer, and A.~Tapp.
\newblock Quantum counting.
\newblock In {\em Proceedings of 25th ICALP}, volume 1443 of {\em Lecture Notes
  in Computer Science}, pages 820--831. Springer, 1998.
\newblock quant-ph/9805082.

\bibitem{BuhrmanCleveWigderson98}
H.~Buhrman, R.~Cleve, and A.~Wigderson.
\newblock Quantum vs.\ classical communication and computation (preliminary
  version).
\newblock In {\em Proceedings of 30th STOC}, pages 63--68, 1998.
\newblock quant-ph/9802040.

\bibitem{cemm:revisited}
R.~Cleve, A.~Ekert, C.~Macchiavello, and M.~Mosca.
\newblock Quantum algorithms revisited.
\newblock In {\em Proceedings of the Royal Society of London}, volume A454,
  pages 339--354, 1998.
\newblock quant-ph/9708016.

\bibitem{coppersmith&rivlin:poly}
D.~Coppersmith and T.~J. Rivlin.
\newblock The growth of polynomials bounded at equally spaced points.
\newblock {\em SIAM Journal on Mathematical Analysis}, 23(4):970--983, 1992.

\bibitem{dodis&khanna}
Y.~Dodis and S.~Khanna.
\newblock Space-time tradeoffs for graph properties.
\newblock In {\em Proceedings of 26th ICALP}, 1999.
\newblock Available at {\tt http://theory.lcs.mit.edu/\~{ }yevgen/} {\tt academic.html}.

\bibitem{ffkl:toolkit}
S.~Fenner, L.~Fortnow, S.~Kurtz, and L.~Li.
\newblock An oracle builder's toolkit.
\newblock In {\em Proceedings of the 8th {IEEE} Structure in Complexity Theory
  Conference}, pages 120--131, 1993.

\bibitem{fortnow&rogers:limitations}
L.~Fortnow and J.~Rogers.
\newblock Complexity limitations on quantum computation.
\newblock In {\em Proceedings of the 13th {IEEE} Conference on Computational
  Complexity}, pages 202--209, 1998.
\newblock cs.CC/9811023.

\bibitem{grover:search}
L.~K. Grover.
\newblock A fast quantum mechanical algorithm for database search.
\newblock In {\em Proceedings of 28th STOC}, pages 212--219, 1996.
\newblock quant-ph/9605043.

\bibitem{grover:framework}
L.~K. Grover.
\newblock A framework for fast quantum mechanical algorithms.
\newblock In {\em Proceedings of 30th STOC}, pages 53--62, 1998.
\newblock quant-ph/9711043.

\bibitem{hajnal:graphprop}
P.~Hajnal.
\newblock An {$n^{4/3}$} lower bound on the randomized complexity of graph
  properties.
\newblock {\em Combinatorica}, 11:131--143, 1991.
\newblock Earlier version in Structures'90.

\bibitem{hh:oneway}
J.~Hartmanis and L.~Hemachandra.
\newblock One-way functions, robustness and the non-isomorphism of
  {NP}-complete sets.
\newblock In {\em Proceedings of the 2nd {IEEE} Structure in Complexity Theory
  Conference}, pages 160--174, 1987.

\bibitem{hkm:qmajority}
T.~Hayes, S.~Kutin, and D.~v. Melkebeek.
\newblock On the quantum complexity of majority.
\newblock Technical Report TR-98-11, University of Chicago, Computer Science
  Department, 1998.

\bibitem{hnw:readonce}
R.~Heiman, I.~Newman, and A.~Wigderson.
\newblock On read-once threshold formulae and their randomized decision tree
  complexity.
\newblock {\em Theoretical Computer Science}, 107(1):63--76, 1993.
\newblock Earlier version in Structures'90.

\bibitem{hromkovic&schnitger:limadvice}
J.~Hromkovic and G.~Schnitger.
\newblock Nondeterministic communication with a limited number of advice bits.
\newblock In {\em Proceedings of 28th STOC}, pages 551--560, 1996.

\bibitem{kss:topo}
J.~Kahn, M.~Saks, and D.~Sturtevant.
\newblock A topological approach to evasiveness.
\newblock {\em Combinatorica}, 4:297--306, 1984.
\newblock Earlier version in FOCS'83.

\bibitem{ks:disj}
B.~Kalyanasundaram and G.~Schnitger.
\newblock The probabilistic communication complexity of set intersection.
\newblock {\em SIAM Journal on Computing}, 5(4):545--557, 1992.

\bibitem{king:graphprop}
V.~King.
\newblock Lower bounds on the complexity of graph properties.
\newblock In {\em Proceedings of 20th STOC}, pages 468--476, 1988.

\bibitem{klauck:qpcom}
H.~Klauck.
\newblock On quantum and probabilistic communication, 
\newblock forthcoming manuscript, 1999.

\bibitem{kushilevitz&nisan:cc}
E.~Kushilevitz and N.~Nisan.
\newblock {\em Communication Complexity}.
\newblock Cambridge University Press, 1997.

\bibitem{lovasz:graphprop}
L.~Lov{\'a}sz and N.~Young.
\newblock Lecture notes on evasiveness of graph properties.
\newblock Technical report, Princeton University, 1994.
\newblock Available at {\tt
  http://www.uni-paderborn.} {\tt de/fachbereich/AG/agmadh/WWW/english/}
{\tt scripts.html}.

\bibitem{minsky&papert:perceptrons}
M.~Minsky and S.~Papert.
\newblock {\em Perceptrons}.
\newblock MIT Press, Cambridge, MA, 1968.
\newblock Second, expanded edition 1988.

\bibitem{mosca:eigen}
M.~Mosca.
\newblock Quantum searching, counting and amplitude amplification by
  eigenvector analysis.
\newblock In {\em MFCS'98 workshop on Randomized Algorithms}, 1998.

\bibitem{nisan:pram&dt}
N.~Nisan.
\newblock {CREW PRAM}s and decision trees.
\newblock {\em SIAM Journal on Computing}, 20(6):999--1007, 1991.
\newblock Earlier version in STOC'89.

\bibitem{nisan&szegedy:degree}
N.~Nisan and M.~Szegedy.
\newblock On the degree of {B}oolean functions as real polynomials.
\newblock {\em Computational Complexity}, 4(4):301--313, 1994.
\newblock Earlier version in STOC'92.

\bibitem{paturi:degree}
R.~Paturi.
\newblock On the degree of polynomials that approximate symmetric {B}oolean
  functions (preliminary version).
\newblock In {\em Proceedings of 24th STOC}, pages 468--474, 1992.

\bibitem{raz:qcc}
R.~Raz.
\newblock Exponential separation of quantum and classical communication
  complexity.
\newblock In {\em Proceedings of 31th STOC}, pages 358--367, 1999.

\bibitem{razborov:disj}
A.~Razborov.
\newblock On the distributional complexity of disjointness.
\newblock {\em Theoretical Computer Science}, 106(2):385--390, 1992.

\bibitem{rivest&vuillemin}
R.~Rivest and S.~Vuillemin.
\newblock On recognizing graph properties from adjacency matrices.
\newblock {\em Theoretical Computer Science}, 3:371--384, 1976.

\bibitem{rivlin:chebyshev}
T.~J. Rivlin.
\newblock {\em Chebyshev Polynomials: From Approximation Theory to Algebra and
  Number Theory}.
\newblock Wiley-Interscience, second edition, 1990.

\bibitem{saks&wigderson:trees}
M.~Saks and A.~Wigderson.
\newblock Probabilistic {B}oolean decision trees and the complexity of
  evaluating game trees.
\newblock In {\em Proceedings of 27th FOCS}, pages 29--38, 1986.

\bibitem{saks&werman:majority}
M.~E. Saks and M.~Werman.
\newblock On computing majority by comparisons.
\newblock {\em Combinatorica}, 11(4):383--387, 1991.

\bibitem{santha:montecarlo}
M.~Santha.
\newblock On the {M}onte {C}arlo decision tree complexity of read-once
  formulae.
\newblock In {\em Proceedings of the 6th {IEEE} Structure in Complexity Theory
  Conference}, pages 180--187, 1991.

\bibitem{snir:trees}
M.~Snir.
\newblock Lower bounds for probabilistic linear decision trees.
\newblock {\em Theoretical Computer Science}, 38:69--82, 1985.

\bibitem{yao:bipartite}
A.~C.-C. Yao.
\newblock Monotone bipartite graph properties are evasive.
\newblock {\em SIAM Journal on Computing}, 17(3):517--520, 1988.

\end{thebibliography}

\appendix

\section{Proof of Theorem~\ref{thqerrorboundt}}

Here we prove a lower bound on small-error quantum search.
The key lemma of~\cite{bbcmw:polynomials} gives the following relation
between a $T$-query network and a polynomial that expresses its acceptance
probability as a function of the input $X$ 
(such a relation is also implicit in some
of the proofs of \cite{fortnow&rogers:limitations,ffkl:toolkit}):

\begin{lemma}\label{lemalgopoly}
The acceptance probability of a quantum network that makes $T$ queries to
a black-box $X$, can be written as a real-valued multilinear $N$-variate
polynomial $P(X)$ of degree at most $2T$.
\end{lemma}

An $N$-variate polynomial $P$ of degree $d$ can be reduced to a single-variate
one in the following way (due to~\cite{minsky&papert:perceptrons}).
Let the {\em symmetrization}
$P^{sym}$ be the average of $P$ over all permutations of its input:
$$
P^{sym}(X)=\frac{\sum_{\pi\in S_N}P(\pi(X))}{N!}.
$$
$P^{sym}$ is an $N$-variate polynomial of degree at most $d$.
It can be shown that there is a single-variate polynomial $Q$ of degree
at most $d$, such that $P^{sym}(X)=Q(|X|)$ for all $X\in\{0,1\}^N$.
Here $|X|$ denotes the Hamming weight (number of 1s) of $X$.

Note that a quantum search algorithm $A$ can be used to 
compute the OR-function of $X$ (i.e.~decide whether $X$ contains at 
least one 1): we let $A$ return some $j$ and then we output the bit $x_j$.
If OR$(X)=0$, then we give the correct answer with certainty; if
OR$(X)=1$ then the probability of error $\e$ is the same as for $A$.
Rather than proving a lower bound on search directly, we will prove a
lower bound on computing the OR-function; this clearly implies 
a lower bound for search.
The main idea of the proof is the following. 
By Lemma~\ref{lemalgopoly}, the acceptance probability of 
a quantum computer with $T$ queries that computes the OR with error probability
$\leq\e$ (under the promise that there are either 0 or at least $t$
solutions) can be written as a multivariate polynomial $P$
of degree $\leq 2T$ of the $N$ bits of $X$.
This polynomial has the properties that
\begin{quote}
$P(\olo)=0$ %
\footnote{\label{noteso}Since we can always test whether we actually found 
a solution at the expense of one more query, we can assume the algorithm 
always gives the right answer `no' if the input contains only 0s.
Hence $s(0)=0$. However, our results remain unaffected up to constant factors 
if we also allow a small error here (i.e.\ $0\leq s(0)\leq\e$).}\\
$1-\e\leq P(X)\leq 1$ whenever $|X|\in[t,N]$
\end{quote}
By symmetrizing, $P$ can be reduced to a single-variate polynomial $s$ 
of degree $d\leq 2T$ with the following properties: 
\begin{quote}
$s(0)=0$\\
$1-\e\leq s(x)\leq 1$ for all integers $x\in[t,N]$
\end{quote}
We will prove a lower bound on $\e$ in terms of $d$. 
Since $d\leq 2T$, this will imply a lower bound on $\e$ 
in terms of $T$. 
Our proof uses three results about polynomials.
The first gives a general bound for polynomials that are bounded by 
1 at integer points~\cite[p.~980]{coppersmith&rivlin:poly}:

\begin{theorem}[Coppersmith \&\ Rivlin]\label{thcopprivlin}
For every polynomial $p$ of degree $d$ that has absolute value
$$
|p(x)|\leq 1 \mbox{ for all integers } x\in[0,n],
$$ 
we have
$$
|p(x)|< a e^{bd^2/n} \mbox{ for all real } x\in[0,n],
$$
where $a,b>0$ are universal constants.
(No explicit values for $a$ and $b$ are given in~\cite{coppersmith&rivlin:poly}.)
\end{theorem}

The second two tools concern the Chebyshev polynomials $T_d$, defined  
as~\cite{rivlin:chebyshev}:
$$
T_d(x)=\frac{1}{2}
\left(\left(x+\sqrt{x^2-1}\right)^d+\left(x-\sqrt{x^2-1}\right)^d\right).
$$
$T_d$ has degree $d$ and its absolute value $|T_d(x)|$ 
is bounded by 1 if $x\in[-1,1]$.
Among all polynomials with those two properties, $T_d$ grows fastest
on the interval $[1,\infty)$
(\cite[p.108]{rivlin:chebyshev} and~\cite[Fact~2]{paturi:degree}):

\begin{theorem}\label{thchebextremal}
If $q$ is a polynomial of degree $d$ such that $|q(x)|\leq 1$ 
for all $x\in[-1,1]$ then $|q(x)|\leq|T_d(x)|$ for all $x\geq 1$.
\end{theorem}

Paturi (\cite[before Fact~2]{paturi:degree} and personal communication) proved

\begin{lemma}[Paturi]\label{lemchebbound}
$T_d(1+\mu)\leq e^{2d\sqrt{2\mu+\mu^2}}$ for all $\mu\geq 0$.
\end{lemma}

\begin{proof}
For $x=1+\mu$:
$\displaystyle T_d(x)\leq (x+\sqrt{x^2-1})^d=(1+\mu+\sqrt{2\mu+\mu^2})^d\leq
(1+2\sqrt{2\mu+\mu^2})^d \leq e^{2d\sqrt{2\mu+\mu^2}}$.
\end{proof}

Now we can prove:

\begin{theorem}\label{thboundepst}
Let $1\leq t<N$ be an integer.
Every polynomial $s$ of degree $d\leq N-t$ such that $s(0)=0$ and
$1-\e\leq s(x)\leq 1$ for all integers $x\in[t,N]$ has
$$
\e\geq\frac{1}{a}e^{-bd^2/(N-t)-4d\sqrt{tN/(N-t)^2}},
$$
where $a,b$ are as in Theorem~\ref{thcopprivlin}.
\end{theorem}

\begin{proof}
A polynomial $p$ with $p(0)=0$ and $p(x)=1$ for all integers $x\in[t,N]$
must have degree $>N-t$.
Since $d\leq N-t$ for our $s$, we have $\e>0$.
Now $p(x)=1-s(N-x)$ has degree $d$ and
\begin{quote}
$0\leq p(x)\leq\e$ for all integers $x\in[0,N-t]$\\
$p(N)=1$
\end{quote}
Applying Theorem~\ref{thcopprivlin} to $p/\e$ 
(which is bounded by 1 at integer points) with $n=N-t$ we obtain:
$$
|p(x)|<\e a e^{bd^2/(N-t)} \mbox{ for all real } x\in[0,N-t].
$$
Now we rescale $p$ to $q(x)=p((x+1)(N-t)/2)$
(i.e.~the domain $[0,N-t]$ is transformed to $[-1,1]$), which has the
following properties:
\begin{quote}
$|q(x)|<\e a e^{bd^2/(N-t)} \mbox{ for all real } x\in[-1,1]$\\
$q(1+\mu)=p(N)=1$ for $\mu=2t/(N-t)$.
\end{quote}
Thus $q$ is ``small'' on all $x\in[-1,1]$ and ``big'' somewhere outside 
this interval ($q(1+\mu)=1$).
Linking this with Theorem~\ref{thchebextremal} and 
Lemma~\ref{lemchebbound} we obtain
\begin{eqnarray*}
1 & = & q(1+\mu)\\
 & \leq & \e a e^{bd^2/(N-t)} |T_d(1+\mu)|\\
 & \leq & \e a e^{bd^2/(N-t)}e^{2d\sqrt{2\mu+\mu^2}}\\
 & = &\e a e^{bd^2/(N-t)+2d\sqrt{4t/(N-t)+4t^2/(N-t)^2}}\\
 & = &\e a e^{bd^2/(N-t)+4d\sqrt{tN/(N-t)^2}}.
\end{eqnarray*}
Rearranging gives the bound.
\end{proof}

Since a quantum search algorithm with $T$ queries induces a polynomial
$s$ with the properties mentioned in Theorem~\ref{thboundepst} 
and $d\leq 2T$, we obtain the following bound for quantum search 
under the promise (if $T\leq N-t$, then $\e>0$):

\bigskip

\noindent
{\bf Theorem~\ref{thqerrorboundt}}\
Under the promise that the number of solutions is at least $t$,
every quantum search algorithm that uses $T\leq N-t$ queries 
has error probability
$$
\e\in\Omega\left(e^{-4bT^2/(N-t)-8T\sqrt{tN/(N-t)^2}}\right).
$$

\section{Proof of Lemma~\ref{lemandord}}

\noindent
{\bf Lemma~\ref{lemandord}}\
Let $d\geq 1$ and let $f$ denote the uniform $d$-level AND-OR tree 
on $N$ variables that has an OR as root. 
There exists a quantum algorithm $A_1$ that finds a 1-certificate 
in expected number of queries $O(N^{1/2+1/2d})$ if $f(X)=1$ and 
does not terminate if $f(X)=0$.
Similarly, there exists a quantum algorithm $A_0$ that finds a 0-certificate 
in expected number of queries $O(N^{1/2+1/d})$ if $f(X)=0$ and 
does not terminate if $f(X)=1$.

\bigskip

\begin{proof}
By induction on $d$.

{\bf Base step.} For $d=1$ the bounds are trivial.

{\bf Induction step (assume the lemma for $d-1$).} 
Let $f$ be the uniform $d$-level AND-OR tree on $N$ variables.
The root is an OR of $N^{1/d}$ subtrees, each of
which has $N^{(d-1)/d}$ variables.

We construct $A_1$ as follows.
First use multi-level Grover-search as 
in~\cite[Theorem~1.15]{BuhrmanCleveWigderson98}
to find a subtree of the root whose value is 1, if there is one. 
This takes $O(N^{1/2}(\log N)^{d-1})$ queries and works with bounded-error.
By the induction hypothesis there exists an algorithm $A'_0$ with expected
number of $O((N^{(d-1)/d})^{1/2+1/(d-1)})=O(N^{1/2+1/2d})$ queries
that finds a 1-certificate for this subtree (note that the subtree 
has an AND as root, so the roles of 0 and 1 are reversed).
If $A'_0$ has not terminated after, say, 10 times its expected number
of queries, then terminate it and start all over with the multi-level
Grover search.
The expected number of queries for one such run is 
$O(N^{1/2}(\log N)^{d-1})+10\cdot O(N^{1/2+1/2d})=O(N^{1/2+1/2d})$.
If $f(X)=1$, then the expected number of runs before success 
is $O(1)$ and $A_1$ will find a 1-certificate after a total expected 
number of $O(N^{1/2+1/2d})$ queries.
If $f(X)=0$, then the subtree found by the multi-level Grover-search
will have value 0, so then $A'_0$ will never terminate by itself and 
$A_1$ will start over again and again but never terminates.

We construct $A_0$ as follows.
By the induction hypothesis there exists an algorithm $A'_1$ with expected
number of $O((N^{(d-1)/d})^{1/2+1/2(d-1)})=O(N^{1/2})$ queries that finds
a 0-certificate for a subtree whose value is 0, and that runs forever if the
subtree has value 1.
$A_0$ first runs $A'_1$ on the first subtree until it terminates, then
on the second subtree, etc.
If $f(X)=0$, then each run of $A'_1$ will eventually terminate
with a 0-certificate for a subtree, and the 0-certificates of the 
$N^{1/d}$ subtrees together form a 0-certificate for $f$.
The total expected number of queries is the sum of the expectations over 
all $N^{1/d}$ subtrees, which is $N^{1/d}\cdot O(N^{1/2})=O(N^{1/2+1/d})$.
If $f(X)=1$, then one of the subtrees has value 1 and the run of $A'_1$
on that subtree will not terminate, so then $A_0$ will not terminate.
\end{proof}

\end{document}